\documentclass[preprint,12pt]{aastex}
\pdfoutput=1 
\begin{document}
\makeatletter
\newenvironment{inlinetable}{%
\def\@captype{table}%
\noindent\begin{minipage}{0.999\linewidth}\begin{center}\footnotesize}
{\end{center}\end{minipage}\smallskip}

\newenvironment{inlinefigure}{%
\def\@captype{figure}%
\noindent\begin{minipage}{0.999\linewidth}\begin{center}}
{\end{center}\end{minipage}\smallskip}
\makeatother

\def\num1{${\cal D}_1$}
\def\mess{${{\theta_E}\over{\theta_1}}$}
\def\tetae{{$\tau_1 ({\rm days})$}}  
\def\br1br{${\cal R}_1^b ({\rm yr}^{-1})$}  
\def\r1d{${\cal R}_1^d ({\rm yr}^{-1})$}  
\def\rtb{{\bf ${\cal R}_{tot}^b ({\rm yr}^{-1})$}}  
\def\rtd{\bf{${\cal R}_{tot}^d ({\rm yr}^{-1})$}}   
\def\gc{globular cluster}
\def\kms{{\rm km/s}}
\def\rmeq#1{\eqno({\rm #1})}
\def\gl{gravitational lens} \def\gb{Galactic bulge}
\def\lc{light curve}
\def\ml{microlensing} \def\mo{monitor}
\def\pr{program} \def\mlmpr{\ml\ \mo\ \pr}
\def\ev{event}
\def\ex{expansion}
\def\fn{function} \def\ch{characteristic}
\def\bi{binary}  \def\bis{binaries}
\def\rd{Di\thinspace Stefano}
\def\bl{binary lens} \def\dn{distribution}
\def\pop{population}
\def\ct{coefficient}
\def\cc{caustic crossing}
\def\cs{caustic structure}
\def\mag{magnification}
\def\pl{point lens}
\def\dm{dark matter}
\def\rd{Di\thinspace Stefano}
\def\tots{track of the source}
\def\de{detection efficiency}
\def\det{detection}
\def\ob{observ}
\def\ol{observational}
\def\od{optical depth}
\def\ml{microlensing}
\def\mtm{monitoring team}
\def\mmtm{microlensing monitoring team}
\def\otm{observing team}
\def\mo{monitor}
\def\motm{microlensing observing team}
\def\los{line of sight}
\def\ev{event}
\def\by{binarity}
\def\ptb{perturb}
\def\sgf{significant}
\def\bis{binaries}
\def\sg{signature}
\def\bbl{binary-lens}
\def\kms{{\rm km/s}}
\def\rmeq#1{\eqno({\rm #1})}
\def\gl{gravitational lens} \def\gb{Galactic bulge}
\def\lc{light curve}
\def\ml{microlensing} \def\mo{monitor}
\def\pr{program} \def\mlmpr{\ml\ \mo\ \pr}
\def\ev{event}
\def\ex{expansion}
\def\fn{function} \def\ch{characteristic}
\def\bi{binary}  \def\bis{binaries}
\def\rd{Di\thinspace Stefano}
\def\bl{binary lens} \def\dn{distribution}
\def\pop{population}
\def\ct{coefficient}
\def\cc{caustic crossing}
\def\cs{caustic structure}
\def\mag{magnification}
\def\ppl{point-lens}
\def\bl{blending}
\def\mage{magnification}
\def\fsse{finite-source-size-effects}
\def\cc{caustic crossing}
\def\mp{multiple peak}
\def\lcf{lightcone fluctuation}
\def\wdf{white dwarf}
\def\pn{planetary nebula}
\vskip -.8 true in
\title{Discovery and Study of Nearby Planetary
and Binary Systems Via Mesolensing}

\author{Rosanne Di\thinspace Stefano}
\affil{Harvard-Smithsonian Center for Astrophysics, 60
Garden Street, Cambridge, MA 02138}

\def\gl{gravitational lensing}
\def\Gl{Gravitational lensing}
\def\ml{microlensing} 
\def\Ml{Microlensing} 
\def\Et{Einstein angle} 
\def\et{$\theta_E$} 
\def\Er{Einstein radius}
\def\ev{event}
\def\vb{variable}
\def\vy{variability}
\def\sg{signature} 
\def\asec{arcsecond}

\begin{abstract}
This paper is devoted to exploring how we can
 discover and study {\it nearby} ($< 1-2$ kpc) planetary and binary 
systems by observing their action as gravitational
lenses.  
Lensing can extend the
realm of nearby binaries and planets that can be systematically studied to
include dark and dim binaries, and face-on systems. 
As with more traditional studies, which use light from the system,
orbital parameters (including the total mass, mass ratio, and
orbital separation) can be extracted from lensing data. 
Also in
common with these traditional methods, individual systems can be
targeted for study. We discuss the specific observing strategies 
needed in order to optimize the discovery and study of nearby
planetary and binary systems by observing their actions as lenses.   

\end{abstract}

\section{Nearby Planets and Binaries}

Microlensing can be a powerful tool to study binaries and planets
\citep{MaoPaczynski, GouldLoeb, DiStefanoMao, DiStefanoScalzo1, 
DiStefanoScalzo2}.
Until recently it has been assumed that most lenses, whether single or
multiple stars, are likely to be at least several kpc away, either in
the Galactic Halo or in the 
distant system of
monitored sources.
It therefore seemed likely that follow-up observations would
be of limited value for multiple systems discovered through
their action as lenses.
In this paper we turn things around and consider lensing by 
binaries and planetary systems that are relatively nearby, within $1-2$ kpc. 
We 
refer to these nearby lenses as {\it mesolenses}
because 
(1) a combination of spatial and time lensing signatures may be
exhibited by the events,  
(2) they
have a high probability, relative to more
distant lenses, of producing detectable lensing events, and
(3) targeted observations designed to study the lensing action of
some selected 
individual masses are possible. (See Di\thinspace Stefano 2005.) 

The lensing science is well-understood. 
 We therefore 
concentrate on contrasts between nearby and more distant multiple lenses,
framing the issues in terms of possible observing programs.  
Section 2 focuses on photometric
monitoring, including ongoing and planned
programs such as Pan-STARRS \citep{PanSTARRS} and 
LSST \citep{LSST}. \S 3 discusses 
ground-based astrometric monitoring, 
which will be carried out by 
both Pan-STARRS and LSST.   
In \S 4 we consider the possibility of selecting individual 
stars to be targets of sensitive lensing studies. 
In \S 5 we discuss what lensing can reveal 
about companion stars and
planets, listing our expectations for the next decade in \S 5.1,
and, in \S 5.2,  emphasizing the
contrasts and interplay with other types of investigations,
such as Doppler and transit studies.   
 
\section{Photometric Monitoring Programs}

Monitoring of large, dense background fields has led to the discovery
of thousands of lensing events. See, e.g.,
\citep{MACHO.Bulge, OGLEBulge, MOA.Bulge, MACHO-98-SMC-1}.
Microlensing monitoring programs have already begun   
to discover nearby ($< 1-2$ kpc) lenses  
\citep{Drake.2004, Nguyen.2004, Kallivayalil.2006}. 
It is only a matter of time
before a nearby planet or binary is discovered through its action
as a lens. This is very likely to happen in data already collected, or being
collected. The OGLE team, e.g., is presently discovering
more than $500$ lensing events per year \citep{OGLEIII.Udalski}. A significant
fraction of these are caused by lenses located within $1-2$ kpc;
most of these lenses are in binaries and/or have planets.
Light curves associated with binary and planet lenses have been 
discovered \citep{OGLE-7, OGLEBinaries1997, OGLEBinaries2002, OGLEBinaries2004,
MACHOBinaries, MACHO-98-SMC-1, OGLE-2003-BLG-235, 
OGLE-2005-BLG-390Lb, OGLE-2005-BLG-169}. 
None of the known multiple lenses appear to be
nearby.\footnote{This is not surprising, since only a handful of lenses
are known to be nearby, and only a few dozen light curves in total 
have been definitely established as caused by binary lenses. 
Also, as Figure 1 demonstrates, the signatures of nearby binaries can
be different.}  
The discovery rate will increase dramatically during the coming decade as  
a new generation of monitoring
programs begins, including Pan-STARRS and LSST. 
Tens of thousands of events caused by nearby lenses alone will be discovered
over a decade (Di\thinspace Stefano 2007). 
Good photometry and astrometry will allow a significant fraction of 
all binary and planet lens events to be identified. 


\noindent{\bf The Einstein Ring:} The image of
a source at a distance of $D_S$, located directly behind a lens
of mass $M$ at a distance $D_L$,
is a ring with angular radius equal to the
Einstein angle, $\theta_E.$
\begin{equation}
\theta_E=.012''\, \sqrt{\Big({{M}\over{M_\odot}}\Big)
                        \Big({{50\, {\rm pc}}\over{D_L}}\Big)
                       \Big(1-{{D_L}\over{D_S}}\Big)}.
\end{equation}
The Einstein radius is $R_E = \theta_E D_L.$ For $M = 1\, M_\odot,$ 
and $D_L = 50$ {\rm pc},  $R_E = 0.6$ AU. This is smaller
than the values typical for microlensing.
Let $\alpha$ be the ratio between the projected orbital
separation, $a,$ and $R_{E}$: $a = \alpha\, R_{E}$.
Binary effects
 are easiest to detect when $0.2 < \alpha < 2$.

\noindent{\bf Rotating Binary Lenses:}
In standard microlensing, typical values of $R_E$  
are several AU. Binarity is detected for systems with
orbital periods on the order of years, generally significantly
longer than the event durations. Model fits either do not
require orbital motion, or require only small phase changes 
during the event  
\citet{Dominik1998, MACHO97-BLG-41-PLANET, EROS-BLG-2000-5}. 
The time scale for photometric events is set by the value of 
$\tau_{E} = 2\, R_{E}/v,$ where $v$ is the transverse velocity.
If the orbital period is $P,$ 
\begin{equation} 
{{\tau_{E}}\over{P}}= {{2.8}}\, 
\Big({{0.2}\over{\alpha}}\Big)^{{3}\over{2}}\, 
    {{50\, {km/s}}\over{v_T}}\,  \Bigg[
{{M}\over{M_\odot}}\, {{50\, {\rm pc}}\over{{D_L}\, (1-x)}}
\Bigg]^{{1}\over{4}}   
\end{equation}
For nearby binary lenses, this ratio 
can be larger than unity.
Figure 1 shows light curves for nearby low-mass binaries completing 
at least one complete orbital cycle during
a lensing event. These light curves,
which happen to have been generated for face-on circular orbits,
 cannot be fit by static models.
At least one additional physical model parameter, the orbital period, is
needed. 
Deriving the value of $P$ provides a relation  
between 
the total lens mass and the value of the semimajor axis.
 
\noindent{\bf Intriguing Binaries with dim components, and rotation 
signatures:} 
The multiples most likely to be discovered by monitoring programs
are those with minimal blending of light from the lens:
 binaries consisting of low-mass dwarfs
or dim stellar remnants are favored. We therefore considered binaries
and planetary systems with primaries that were black holes (BHs), neutron
stars (NSs), and low-mass dwarfs. 
We generated a large number of binaries and planetary systems; the models are
described in the caption to Figure 2. 
We selected those for which evidence of binarity
is potentially detectable ($\alpha$ in the range $0.2-2$), and
for which significant rotation would be expected during a lensing event
(we required rotation through $2\, \pi$ while the source was within
$2\, \theta_E$ of the lens). 
The average separations are shown in Figure 2.
For BH binaries, the average orbital separation  
lies between a few tenths of an AU through a few AU.
This range includes the natural end states of those mass transfer binaries with
a high-mass (low-mass) donor which eventually becomes a neutron star
(white dwarf). Binaries of this type should exist and can be
discovered through lensing, as can binaries in which a BH is
orbited by a dwarf star or planet. Similar statements apply to
NSs, in that the range of orbital separations
shown in Figure 2 encompasses interesting classes of double-degenerate
and NS/low-mass-companion systems.
   If the primary is a low-mass dwarf 
(bottom two sets of lines), then both components have low mass.
This regime is also unexplored. Interestingly enough, some of the
close M-dwarf/planet separations satisfying our conditions,
would place the planet in the habitable zone (Di~Stefano \& Night 2007).

\section{Monitoring for Astrometric Events}

Lensing produces centroid shifts, which have been well-studied for
point lenses (Dominik \& Sahu 2000; Bozza 2001; An \& Han 2002). Let 
$u$ be the angle of closest approach between the lens and source,
expressed in units of $\theta_E.$  For $u<<1,$ 
the angular shifts in the position of bright images
 tend to be in the range of tenths of $\theta_E;$
the effect decreases as $1/u.$ This has the implication that
monitoring programs sensitive to centroid shifts larger than a few
milliarcseconds will be able to detect the astrometric effects caused by  
nearby lenses
for values of $u$ that can be larger than unity.   
Monitoring conducted by
programs like Pan-STARRS and LSST will have the required 
sensitivity.\footnote{In fact these missions will 
measure the
parallax and proper motions of a large fraction of the
 stars within roughly $200$ pc.
These shifts are comparable to those expected in many 
mesolensing events.}

For point lenses, the
centroid of light traces an ellipse which starts and ends at the
true source position.  
For this paper, the most relevant question is whether the shifts
produced by binary lenses are sufficiently
different, to allow us
to determine that the lens is a binary.
Figure 3 demonstrates that binary lenses can 
produce deviations from this pattern with sizes in the
range $0.1 -1 \, \theta_E.$ Therefore, depending on the value of
$\theta_E,$ the binary parameters, 
and the path of the source behind the lens,
the spatial deviations associated with binaries can be
detected and measured.
The motion of the centroid encodes information about the mass of the lens
and its binary characteristics (mass ratio and projected orbital
separation) that can be used to determine the properties of the lens.

\section{Targeted Observations: lensing as follow-up}

Let ${\cal R}_1$ represent the average rate at which a single lens 
generates detectable events. Because ${\cal R}_1$ is proportional
to $D_L^{-1.5},$ its value for nearby lenses can be high. A solar-mass
star with transverse velocity $v=50$ km s$^{-1}$, and $D_L = 50$ pc,
produces events at an average rate of $0.05$ per year, provided that
the average stellar density of the background is greater than roughly
a few per sq.\, arcsec. Details are presented in Di Stefano (2005).
We can take the rate for an individual point lens 
to be a lower bound,  
since
lens multiplicity increases the rate by a factor that
approaches 
$\sqrt{1+q}$ as the value of $a$ becomes larger than roughly $2\, R_E.$   

Given 100 nearby lenses in front of dense source fields,   
we therefore expect, on average, several detectable events per year.  
(Note that, per sq.\, degree, there are $\sim 100$ dwarf stars located within 
$200$ pc.)
By examining the backgrounds behind potential lenses, 
we can identify those in front of particularly promising regions 
(e.g., those in which the lens proper motion is
directed toward a star or an upward fluctuation in the 
surface brightness).     
In addition, detectability can be enhanced by more sensitive
observations. The rate quoted above assumed ten percent photometry.
If, instead, our observations can reliably detect a $1\%$ deviation
from baseline, the rate will be $10$ times higher. (${\cal R}_1$ 
is proportional to $f_T^{-1},$ where $f_T$ is the fractional increase in the 
baseline light needed for a detection.)  
It is also important that ${\cal R}_1$
is proportional to $\theta^{-2},$ where $\theta$ is the smallest region
over which we can reliably detect a deviation from baseline; this 
is because the lensing of a single star is more easily detected
if the total number of stars per resolution element is small. 
For a large variety of 
dense backgrounds, and under optimal observing conditions,
an individual lens can produce events at a sequence of times
that can be predicted (Di Stefano 2005). In some cases, 
event detection will
require masking light from the primary star. 
For a discussion of the  history of proposed targeted studies
see Di\thinspace Stefano 2005, \S 3.2.4.   
Figures 1, 3, and
 4 illustrate
what can be learned about binaries and planets
by studying events, or even sequence of events.

\section{What we can learn from lensing}

\subsection{Expectations} 

\noindent (1) Monitoring programs will discover and correctly identify
 planetary systems and binaries
located within a kiloparsec, many exhibiting significant orbital motion. 
The components of these nearby multiples
will include stellar remnants, low-mass stars, brown dwarfs, and planets.
If $10\%$ of the lenses are nearby, and $10\%$ ($1\%$) of binary (planet)
lenses can be identified, then ten years of OGLE will identify dozens
of nearby binaries and a handful of planets. 
If future observing programs can detect 
deviations from baseline of $\sim 2\%$, against dense stellar  
regions $\sim 0.5''$ on a side, 
 the discovery rate of events per field  
will increase by a factor of $20.$ In addition,
considering dense backgrounds and
the backgrounds provided by individual stars, 
all-sky monitoring will effectively
cover more than $3$ times the area now regularly monitored 
by OGLE.

\noindent (2) Nearby multiple systems discovered via lensing will be targeted
for multiwavelength studies. If the components are stellar
remnants, follow-up observations at X-ray wavelengths may 
establish their nature. 
Of particular interest 
are stellar remnants in binaries that have completed their
mass transfer histories, but which are not yet copious emitters
of gravity waves.
If the components are cool dwarfs,
spectroscopic studies may measure orbital parameters.   

\noindent (3) Nearby masses in front of distant
dense source fields will be targeted for lensing studies.   
Within $200$ pc there are $\sim 100$ dwarf stars (M or later) 
per sq.\, degree. If $200$ sq.\, degree of the sky provides a
useful background, then lensing by up to $2E4$ very nearby
dwarf
stars, many of which are likely to be multiples, are 
potential targets. 
We must develop ways to systematically
identify those with backgrounds most likely
to produce detectable effects, and to predict the times at which
observations are most likely to be fruitful.    
Even {\sl luminous} stars will be targeted for lensing studies,
by
conducting  
observations in wavebands in which the
potential lens is dark, or employ masking techniques.
Work  
is needed to determine the best ways to block light from the 
primary in a lens to optimize the study of light from the images of
background sources.

\noindent (4) Lensing observations will determine
system parameters for many binaries. 
Under ideal circumstances, the total mass of the
lens system can be determined, as can the orbital period, separation,
and mass ratio.  
Lensing studies can sometimes 
even provide two ways to determine the same
quantity, allowing consistency checks. 

\subsection{Comparisons with other methods}

Doppler shifts and eclipses (or transits) have provided most of the 
information we have about multiple systems. 
Doppler techniques require that enough photons be received from the
system to provide
a well-delineated spectrum; transits require enough photons to
allow us to distinguish small changes in the amount of light received.
Lensing, on the other hand,
 can detect the multiplicity of objects that provide little or
no flux. It can thereby extend the reach of our studies   
to dark and dim objects, such as nearby binaries consisting of
neutron stars and dim low-mass dwarfs with planets. 

To apply Doppler techniques, the velocity of the central star
must have a significant radial component. Transits can only 
be observed if the orbital inclination lies in a relatively narrow,
nearly edge-on 
range. Lensing, in contrast,
 requires only that the components of the multiple are 
separated by $> 0.2\, \theta_E$.   
 Lensing can therefore detect binarity and discover
planets in systems not accessible to other methods.

Lensing studies do, however, 
face some unique challenges.
First, we know that the binary identification of the early monitoring
projects is incomplete   
\citep{DiStefano2000, Night.2007}. 
This, however, is not due to intrinsic barriers,
and is not unique to nearby binaries and planets. In this paper
we therefore 
assume that efforts to    
improve the selection criteria used by new programs will be successful. 
The new challenges special to the study of nearby multiples 
are: (1) establishing the nature of the primary in those cases in which
 the
binary does not emit much radiation. 
(2) developing useful
criteria to decide which known foreground objects, including
luminous stars, are good targets
for lensing studies.

\noindent{\bf Summary:} Until now, lensing has generally been 
viewed as a way to
explore binary and planet populations of distant populations
(see. e.g., Di Stefano 2001). 
We find that lensing can also be used to discover 
nearby multiple systems with dark or dim components. 
Whether the primary is dim or bright, lensing can, in
principle,  measure orbital parameters.
Further work focused on these possibilities will allow lensing to
join Doppler studies
and transit studies as an important and productive 
way to study the population and
properties of nearby binaries and planets.  

\bigskip{\sl Acknowledgements:} I thank
Christopher Night, Florian Dubath, Eric Pfahl, and
Penny Sackett for discussions.         
This work was supported in part by NAG5-10705
and in part by the National Science Foundation under Grant No. PHY99-07949.


\clearpage

\begin{figure}
\epsscale{.7}
\plotone{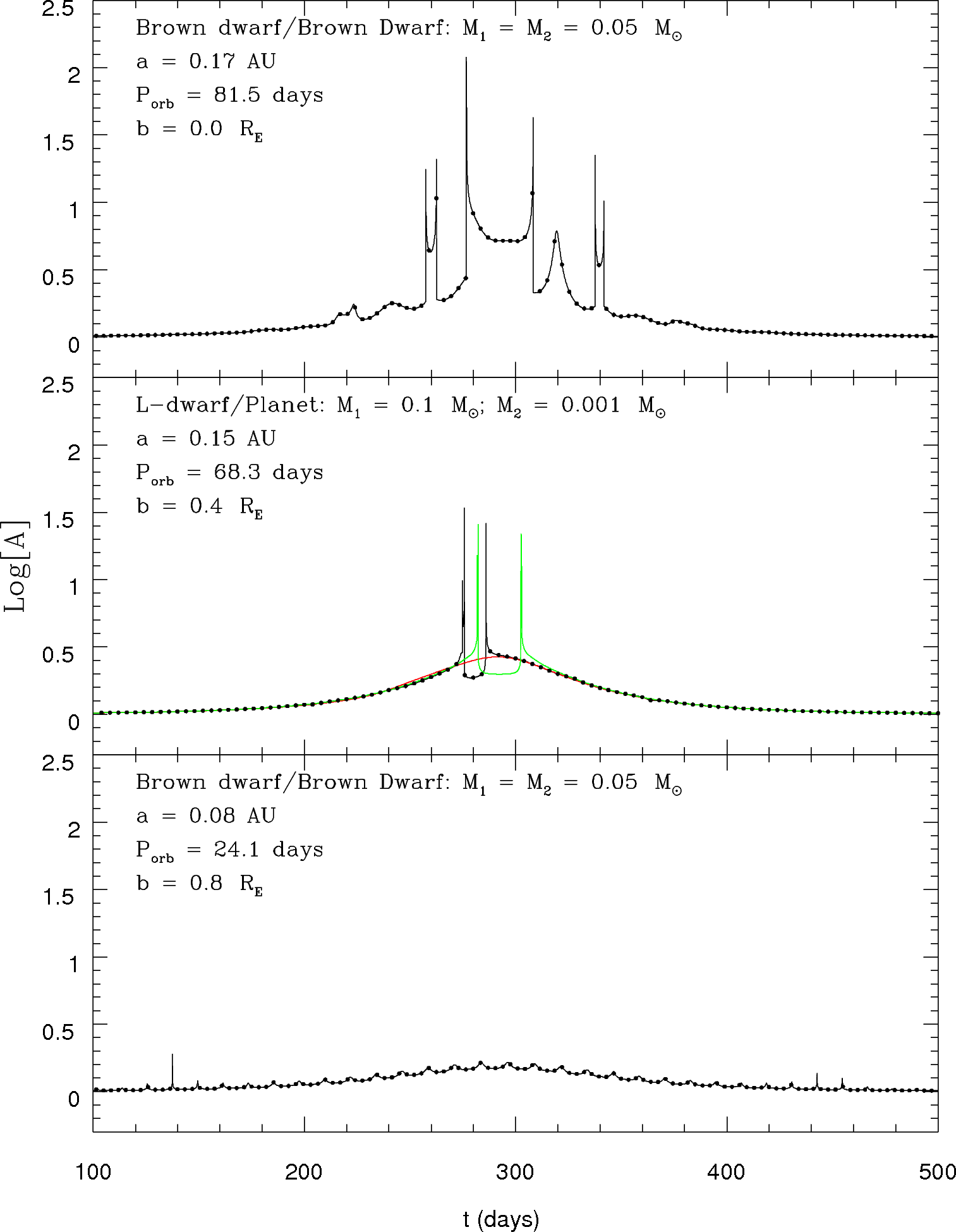}
\vspace{.75 true in}
\caption{\small 
Simulated light curves with more than one orbit occurring during the
lensing event. 
Solid curves: model, with points 
placed at 3.5-4 day intervals to show times of possible monitoring.
{\bf Top} curve exhibits periodic fluctuations in the wings, 
and more peaks than possible in the static case. 
{\bf Middle:} 
Red (green) 
curve is for a static lens oriented along
(perpendicular to) the direction of motion. Comparing model with the
red curve shows that rotation guarantees 
caustic crossings during the event. Comparing with the
green curve shows that orbital motion
decreases the time between characteristic features.
{\bf Bottom:} Regular monitoring would detect a jagged low-amplitude deviation
from baseline; periodicity could be analyzed with Fourier methods.
}
\end{figure}

\begin{figure}
\vspace{-.2 true in}
\plotone{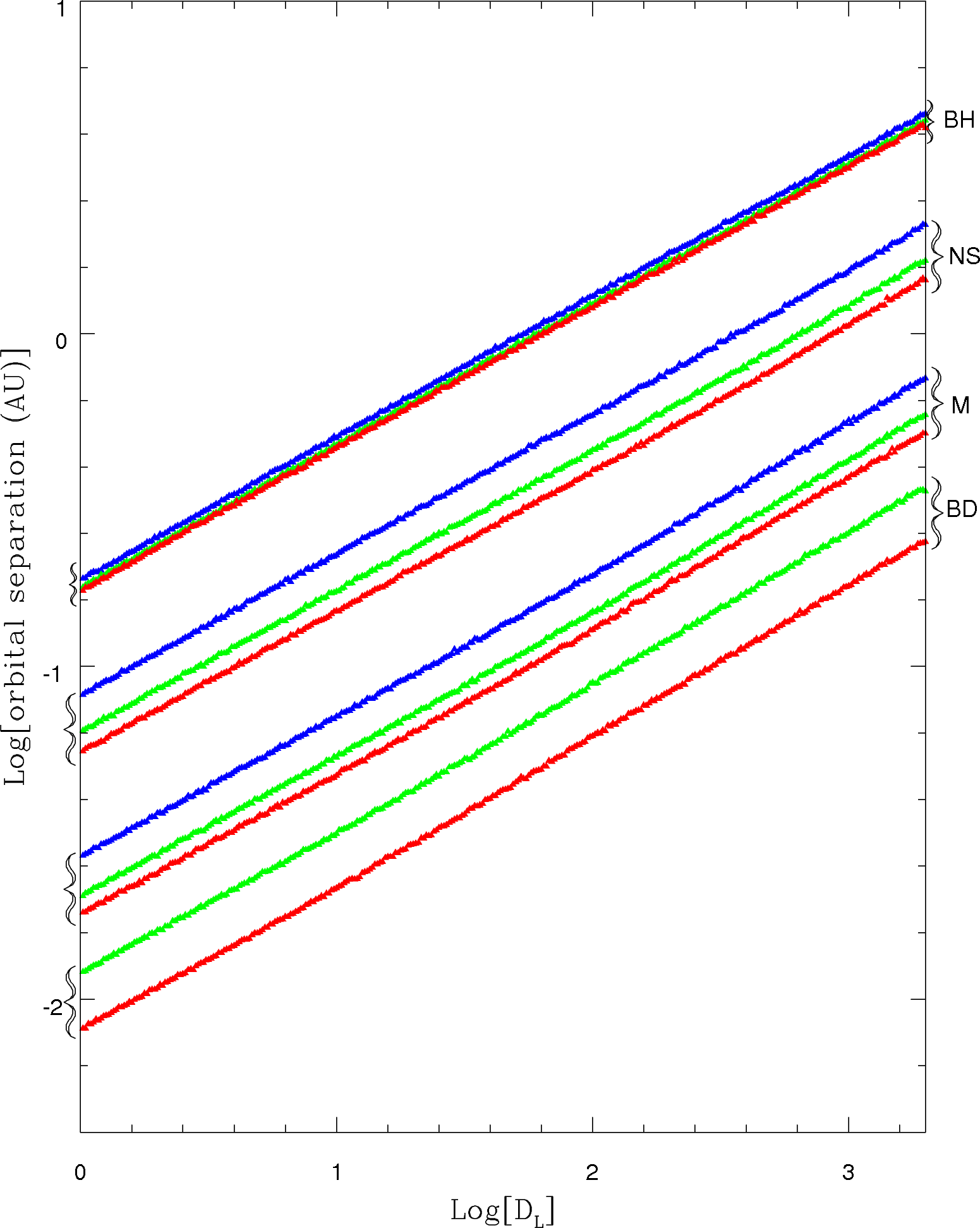}
\epsscale{.60}
\vspace{.7 true in}
\caption{The logarithm of the average orbital separation, in AU, vs
$D_L$  for planets and binary lenses producing light curves
from which $P$ is potentially measurable.
Each system 
has $0.2 < \alpha < 2.$ and 
$\tau_E/P > 0.5$. When a {\bf BH} ($M = 10\, M_\odot$) or  
{\bf NS} ($M = 1.4\, M_\odot$) is the primary, 
the companion is a NS (blue, top curve),
white dwarf ($M = 0.4\, M_\odot$; green, middle curve), or 
planet ($M = 10\, M_{Jupiter}$; red, bottom curve). When the primary
is an 
{\bf M-dwarf} ($M = 0.2\, M_\odot$), the companion is another M-dwarf (blue),
brown dwarf ($M = 0.05\, M_\odot$; green), 
or Jupiter-mass planet (red). When the primary
is a {\bf BD}, the companion is another BD (green)
or Jupiter-mass planet (red).   
}

\end{figure}

\begin{figure}
\plotone{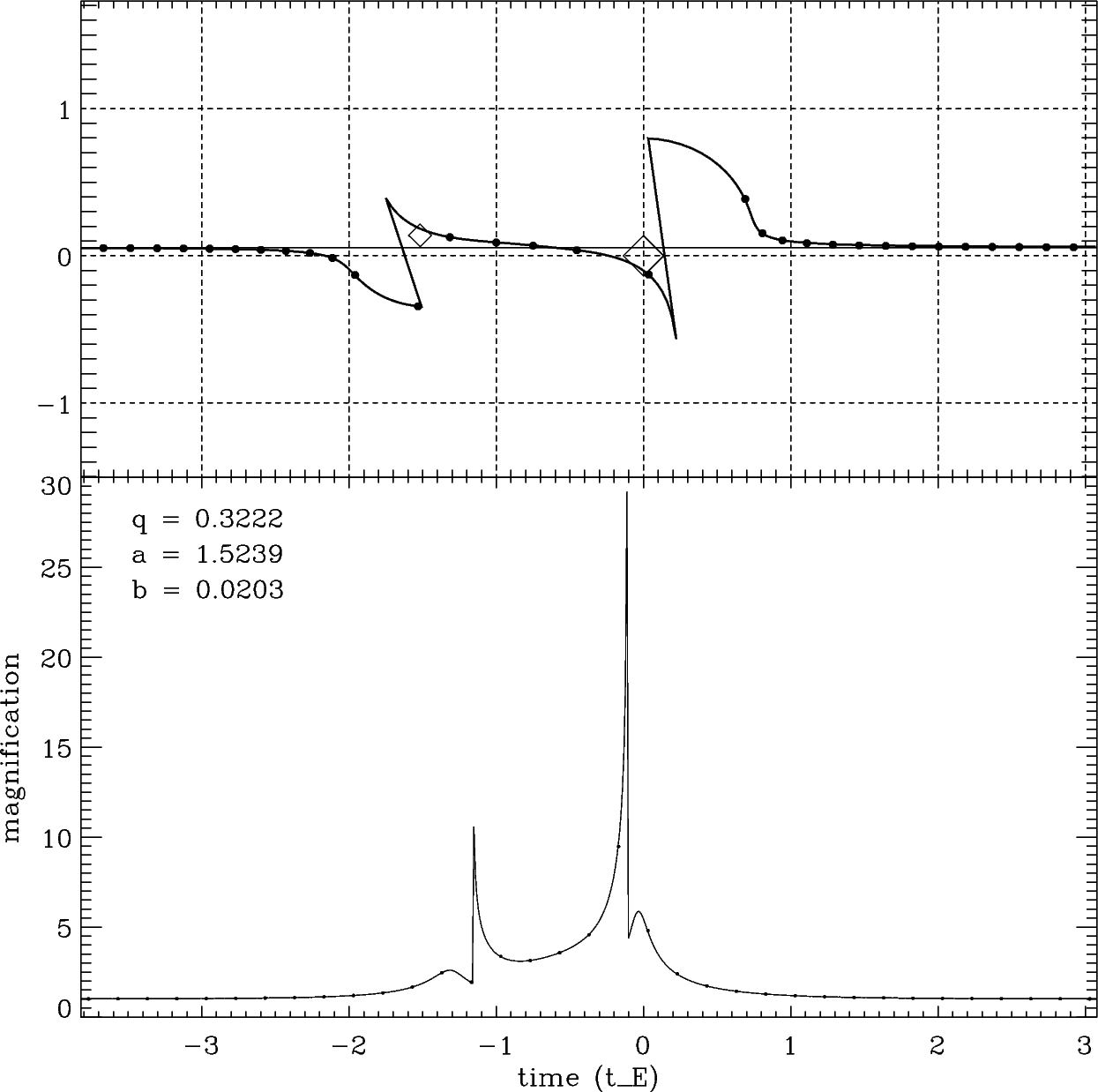}
\caption{
{\bf Top:} Instantaneous coordinates of the centroid position (in units of
$\theta_E$). Diamonds mark the positions of the masses. The trajectory is shown:
it starts on the left; each dark point on the trajectory matches a (lighter) point
on the light curve. {\bf Bottom:} magnification versus the lens position.
Measurable shifts in the centroid position can start prior to and end
after measurable photometric deviations. Centroid motions can be linked
to the size of the Einstein ring; when $D_L$ is known, this becomes
a measurement of the lens mass. Binary separation and mass ratio can
be directly determined for wide binaries. In the light curve
above, the separation between caustics can be determined. 
}
\end{figure}

\begin{figure}
\plotone{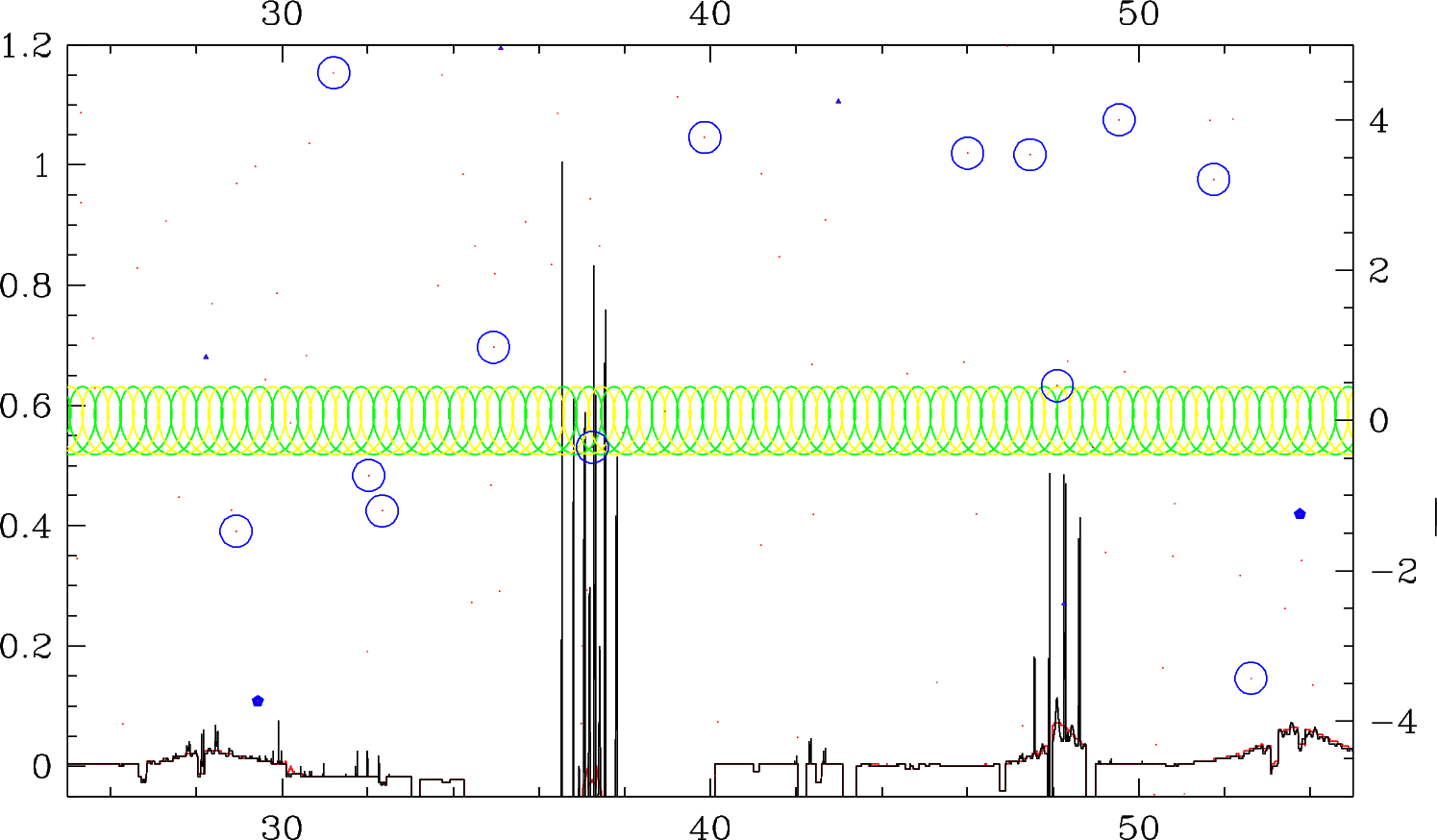}
\caption{A binary lens with 
total mass $0.8\, M_\odot$, with $D_L =11$ pc, and  
$q =1$ and  $a = 0.9\, R_E$ is moving in front of
a field with total stellar density $0.3\, \theta_E^{-2}$, modeled with 
a realistic
 luminosity function. The positions of background stars are marked with,
in order of brightness, red dots, blue hexagons, and open blue circles.  
Only when the binary passes in front of bright stars (in this case, 
$L > 100-1000\, L_\odot$) are lensing effects 
detectable. Horizontal axes: time (x-coordinate) 
in units of the time required to cross $\theta_E$;
the total time shown is roughly 5 years.
Left vertical axis: logarithm of the total magnification (arbitrary
units) of a region of unresolved stars  
of area $25 \theta_E^2$; values below zero correspond to regions with 
baseline luminosity below
the average value.  Right vertical axis: y-coordinate, 
in units of the Einstein crossing time. Two ``events''
separated by almost two years; each event exhibits more structure
because of the binary rotation. Even if it is not 
as well-resolved as in this example,
the additional structure provides 
more information about the binary parameters and translational
motion of the lens.
In this case, the transverse speed is
lower than typical, allowing the details to be well-resolved
with hourly monitoring. 
}
\end{figure}

\end{document}